\documentclass[aps,prl,showpacs,preprint]{revtex4}
\usepackage{graphicx}
\usepackage{amsmath}

\usepackage{epsfig}

\begin{document}

\title{Global Stationary Phase and the Sign Problem}

\author{Andr{\'e} G.\ Moreira,$^1$ Stephan A.\ Baeurle$^2$ and Glenn H.\ Fredrickson$^1$}
\affiliation{$^1$Materials Research Laboratory, University of California, Santa Barbara, CA 93106, USA}
\affiliation{$^2$Institut f{\"u}r Physikalische und Theoretische Chemie, Universit{\"a}t Regensburg, 93053 Regensburg, Germany}

\date{\today}

\begin{abstract}
We present a computational strategy for reducing the sign problem
in the evaluation of high dimensional integrals with non-positive
definite weights. The method  involves stochastic sampling with a
positive semidefinite weight that is adaptively and optimally
determined during the course of a simulation. The optimal
criterion, which follows from a variational principle for analytic
actions $S(z)$, is a global stationary phase condition that the
average gradient of the phase $\mathrm{Im} S$ along the sampling
path vanishes. Numerical results are presented from simulations of
a model adapted from statistical field theories of classical
fluids.
\end{abstract}

\pacs{05.10.-a,02.70.-c,82.20.Wt}
%05.20.-y Classical statistical mechanics

\maketitle

A familiar problem that arises in the context of lattice gauge
theory\cite{lgt}, quantum
chemistry\cite{quantum1}, correlated electron
physics\cite{loh}, and equilibrium field theories of classical
fluids\cite{ghfrev}, is the evaluation of integrals of the
form
\begin{equation}\label{eq1}
  Z=\int_{C_1} dx \; \exp [-S(x)]
\end{equation}
where the path of integration $C_1$ is the real axis and the
action (or effective Hamiltonian) $S(x)$ is \textit{complex}. In
the cases of primary interest $x \in R^n$ is a n-vector
representing a discrete representation (lattice sites or spectral
elements) of one or more classical or quantum fields. The
dimension $n$ is typically large, of order $10^3 - 10^6$. Here we
shall use one-dimensional notation, although the formalism is
primarily intended for cases of $n \gg 1$.

For real $S(x)$, there are a variety of powerful methods available for
evaluating $Z$, including Monte Carlo and (real) Langevin
simulations\cite{binder}. However, in the case of complex
$S=S_R+iS_I$, the integrand is not positive semidefinite, so the Monte
Carlo method is not immediately applicable. Simulations can be carried
out using the positive semidefinite weight $\exp (-S_R )$, but then an
oscillatory phase factor of $\exp (-i S_I )$ must be included in the
computation of averages\cite{lin}. The rapid oscillations in this
factor (the ``sign problem''), which become more pronounced for large
$n$, can dramatically slow convergence in such simulations.
Alternatively, a ``complex Langevin'' simulation technique has been
devised in which the field variables $x$ are extended to the complex
plane and a Langevin trajectory prescribed for the purpose of
generating Markov chains of states\cite{parisi}. Unfortunately this
method is not guaranteed to converge and pathological behavior has
been noted for specific models\cite{lee,schoenmaker}. In the present
letter we describe a new simulation approach that is useful for
reducing the sign problem in integrals of the form of Eq.~(\ref{eq1}),
where $S(z)$ is an \textit{analytic} function of the complex n-vector
$z=x+iy$.

We begin by considering a displacement of the original integration
path along the real $x$ axis, $C_1$ to a new \textit{parallel
path} $C_y$ defined by $z=x+iy, \; x_j \in (-\infty , \infty )$,
in which $y \in R^n$ is an arbitrary displacement of $C_1$ along
the imaginary axis. Note that the displacement $y_j$ need not be
uniform in $j$ for the $n>1$ case. Provided $S(z)$ is analytic in
the rectangular strip bounded by $C_1$ and $C_y$ and $| \exp
[-S(R+iy)] | \rightarrow 0$ for $R \rightarrow \pm \infty$, it
follows that
\begin{equation}\label{eq2}
  Z=\int_{C_y} dz \; \exp [-S(z)] = \int_{C_y} dx \; \exp
  [-S(x+iy)]
\end{equation}
and  the resulting $Z$ is independent of the choice of $y$. Upon
decomposing $S$ into real and imaginary parts $S_R (x,y) +i S_I
(x,y)$, $Z$ can be rewritten as
\begin{equation}\label{eq3}
  Z=Z_y \int_{C_y} dx \; P_y (x) \exp [-i S_I (x,y)]
\end{equation}
where $Z_y \equiv \int_{C_y} dx \; \exp [-S_R (x,y)]$ and $P_y
(x)$ is a normalized, positive semidefinite, probability
distribution for a random variable $x$ at the fixed value of $y$:
\begin{equation}\label{eq4}
  P_y (x) = \exp [-S_R (x,y)]/Z_y
\end{equation}
It follows that the average of an analytic observable $f(x)$ can
be evaluated alternatively from the formulas
\begin{eqnarray}
\langle f(x) \rangle & = & Z^{-1} \int_{C_1} dx \; \exp [-S(x)]
f(x) \nonumber \\
& = & \frac{\langle \exp [- i S_I (x,y)] f(x+iy)
\rangle_y}{\langle \exp [- i S_I (x,y)] \rangle_y} \label{eq5}
\end{eqnarray}
where $\langle h(x) \rangle_y \equiv \int_{C_y} dx \; P_y (x)
h(x)$ denotes an average with probability weight $P_y (x)$.

It is the second expression in Eq.~(\ref{eq5}) that is of interest in
the present letter. A poor choice of $y$ will lead to significant
oscillations in the phase factor $\exp [-i S_I (x,y)]$ as $x$ is
stochastically varied along the sampling path $C_y$ in a simulation.
This would drive both numerator and denominator in Eq.~(\ref{eq5}) to
zero and dramatically slow or prevent convergence of average
quantities of interest. One approach to alleviate this difficulty
would be to choose $y=y^*$, where $y^*$ is the imaginary component of
a saddle point $z^*$ defined by $S^\prime (z^* )=0$. The deformed
integration path $C_{y^*}$ would then be a line passing through the
saddle point parallel to the real axis.  If this path happened to be a
constant phase (steepest ascent) path locally around the saddle point,
then the phase oscillations would be reduced on trajectories that
remain close to the saddle point\cite{bender}. In general, however,
path $C_{y^*}$ will not be a a constant phase path, even in the close
vicinity of $z^*$. A local analysis about each saddle point, costing
$O(n^2 )$ in computational effort, can be used to identify proper
constant phase paths. However, in typical problems where field
fluctuations are strong, significant weight is given to trajectories
that are \textit{not localized} around saddle points.

The essence of our method is a \textit{global} strategy for
selecting an optimal displacement $y$, denoted $\bar{y}$. To this
end, we introduce a ``generating'' function (functional)
\begin{equation}\label{eq6}
  G(y) = \ln \int_{C_y} dx \; \exp [ - S_R (x,y)]
\end{equation}
Invoking the Cauchy-Riemann (CR) equations, it is straightforward
to show that the first derivative of $G(y)$ is given by
\begin{equation}\label{eq7}
  \frac{\partial G(y)}{\partial y_j} = \langle \frac{\partial}{\partial x_j} S_I (x,y)
  \rangle_y
\end{equation}
The second derivative follows from repeated application of the CR
equations and an integration by parts
\begin{eqnarray}
\frac{\partial^2 G(y)}{\partial y_j \partial y_k} & = & \langle [
\frac{\partial}{\partial x_j} S_I - \langle
\frac{\partial}{\partial x_j} S_I \rangle_y ] [
\frac{\partial}{\partial x_k} S_I - \langle
\frac{\partial}{\partial x_k} S_I \rangle_y ] \rangle_y
\nonumber \\
& + & \langle [ \frac{\partial}{\partial x_j} S_R ] [
\frac{\partial}{\partial x_k} S_R ]\rangle_y , \label{eq8}
\end{eqnarray}
which is the sum of two positive definite forms. It follows that
$G(y)$ is manifestly a \textit{convex} function for any $y$.

We now claim that the ``optimal'' choice $y=\bar{y}$ is such that
\begin{equation}\label{eq9}
  \frac{\partial G(y)}{\partial y_j} |_{\bar{y}} =
  \langle \frac{\partial}{\partial x_j} S_I (x,\bar{y})
  \rangle_{\bar{y}} = 0
\end{equation}
Evidently such a point would be a \textit{local minimum} of
$G(y)$. Moreover, it implies that $S_I$ \textit{has vanishing
gradients on average} along the sampling path $C_{\bar{y}}$. This
condition can be viewed as a \textit{global}, rather than
local\cite{bender}, stationary phase criterion and would seem to
be an excellent way to minimize the effect of phase fluctuations.
Since $G(y)$ has a unique minimum, it follows that $\bar{y}_j$ is
homogeneous in $j$ for bulk systems with translationally invariant
actions. The method evidently produces nontrivial inhomogeneous
$\bar{y}$ when applied to field theories in bounded geometries.

It remains to discuss how to incorporate this optimal choice of
sampling path into a simulation algorithm. We propose the
following ``optimal path sampling'' (OPS) algorithm:
\begin{enumerate}
\item
Initialize vectors $x$ and $y=y^k$ with $k=0$.
\item
Carry out a stochastic simulation in $x$ at fixed $y^k$ to
generate a Markov chain of $x$ states of length $M$. $P_{y^k} (x)$
should be used as a statistical weight for importance sampling.
The simulation method could be Metropolis Monte Carlo, its
``smart'' or ``hybrid'' variants\cite{kennedy}, or a real
Langevin technique.
\item
Evaluate $G(y^k )$ and $\partial G (y^k )/\partial y^k$ by
averaging over the $x$ configurations accumulated in the $M$-state
simulation. Update $y$ to approach $\bar{y}$ by making a steepest
descent step
$$
y^{k+1} = y^k - \lambda \frac{\partial G (y^k )}{\partial y^k}
$$
where $\lambda$ is an adjustable relaxation parameter.
Alternatively, the accumulated information on $G(y)$ could be used
to carry out approximate line minimizations, which would permit
conjugate gradient updates from $y^k$ to $y^{k+1}$.
\item
Repeat steps 2 and 3 for $k=1,2, ...$ until the sequence of $y^k$
converges to within some prescribed tolerance to $\bar{y}$. The
simulation has now \textit{equilibrated}.
\item
Carry out a long stochastic simulation (``production run'') with
statistical weight $P_{\bar{y}} (x)$.
\item Compute averages over the simulated states according to
  Eq.~(\ref{eq5}) with $y=\bar{y}$.
\end{enumerate}
Evidently, the parameters $M$ and $\lambda$ can be adjusted to
accelerate the ``equilibration'' period.

\begin{figure}
  \begin{center}
    \epsfig{file=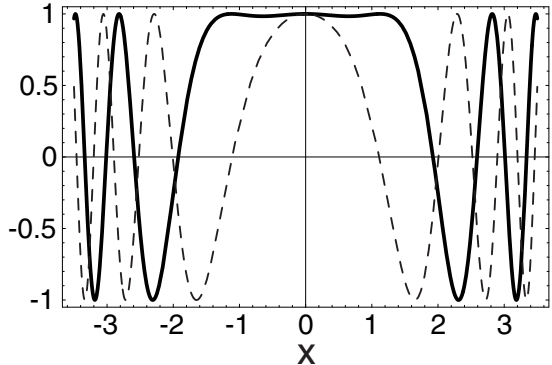,width=7.2cm}
  \end{center}
\caption{Variation of the phase factor of the Airy integrand
$\mathrm{Re}[\exp (-i S_I (x,y,1))]$ with $x$ for $y=0$ (dashed)
and $y=\bar{y}=1.19149$ (solid).} \label{fig:fig1}
\end{figure}

Our OPS method has some similarities to (and was inspired by) the
complex Langevin (CL) simulation technique. In that approach, one
generates a Markov chain of states in the complex plane by
integrating the Langevin equations\cite{parisi}
\begin{equation}\label{eq10}
  \frac{\partial x}{\partial t} = - \mathrm{Re} \frac{d S}{dz}
  +\eta (t)
\end{equation}
\begin{equation}\label{eq11}
  \frac{\partial y}{\partial t} = - \mathrm{Im} \frac{d S}{dz}
\end{equation}
where $\eta (t)$ is a real Gaussian white noise with $\langle \eta
(t) \rangle =0$ and $\langle \eta_j (t) \eta_k (t^\prime ) \rangle
=2 \delta (t-t^\prime )\delta_{jk}$. Ensemble averages $\langle
f(x) \rangle$ are computed as time averages of $f(x+iy)$ over the
chain of states. Under conditions where the CL method converges,
we have observed that $y$ drifts to a nearly constant value that
is not associated with any saddle point $y^*$.  Eq.~(\ref{eq11})
reduces approximately in this case to $\langle \mathrm{Im}\; dS/dz
\rangle_{y} =0$, which is equivalent to the condition (\ref{eq9}).
The OPS technique is also distinct from so-called ``stationary
phase Monte Carlo'' methods, which apply filtering and sparse
sampling methods to suppress phase
oscillations\cite{quantum1,sabo}. These methods
are effective but apparently have no variational basis.

Before providing a numerical example of the OPS method, it is
illustrative to see how our global stationary phase criterion
works in a simple one dimensional example
\begin{equation}\label{11b}
  \mathrm{Ai} (t) = \frac{1}{2 \pi} \int_{-\infty}^\infty dx \;
  \exp[i(x^3 /3 + t x)],
\end{equation}
which is a representation of the Airy function. In this case
$S(x,t) = - i (x^3 /3 + t x)$ and Eq.~(\ref{eq9}) leads to
$\bar{y}^2 - \langle x^2 \rangle_{\bar{y}} - t =0$. This equation
has a single root, corresponding to the minimum of $G(y)$, that
yields $\bar{y} (t)$. For example, $\bar{y} (1)=1.19149$. Of
particular interest is the effect of the optimal displacement on
phase oscillations. In Fig.~\ref{fig:fig1} we plot
$\mathrm{Re}[\exp (-i S_I (x,y,t))]$ verses $x$ at $t=1$ for $y=0$
(no shift) and $y=\bar{y}$ (optimal). Clearly the optimal shift
dramatically suppresses phase oscillations over the interval $-2
\lesssim x \lesssim 2$. The global stationary phase criterion has
no effect outside this interval, because $P_{\bar{y}} (x)$ decays
supra-exponentially there as $\sim \exp (-x^2 \bar{y})$ and so no
statistical weight is given to $|x| \gtrsim 2$.

\begin{figure}
  \begin{center}
    \epsfig{file=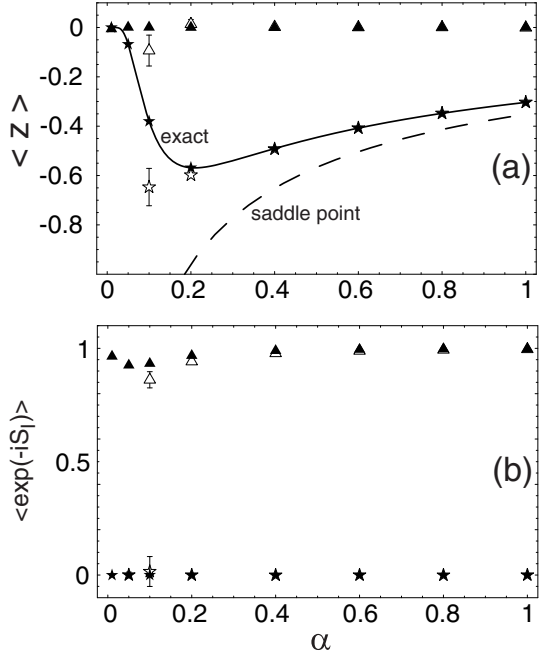,width=7.2cm}
  \end{center}
\caption{Comparison between OPS and CL simulations.
  (a): The average of $z=x+i y$ as a function of the
  parameter $\alpha$ for the model of Eq.~(\ref{eq12}) with $n=1$
  and $\chi=1$. Open and filled symbols are results,
  respectively, from CL and OPS. Stars
  denote the average of the imaginary part $y$ and triangles the average of
  the real part $x$. The full line is the exact solution, the
  dashed line the corresponding saddle point. (b): The
  average sign $\exp (-i S_I )$ for the same parameters as in (a). The convention for
  the symbols is the same as in (a). Error bars are
  comparable to the symbol sizes if not explicitly shown.}
\label{fig:fig2}
\end{figure}

As a numerical test of the OPS method, we have carried out
simulations of the model
\begin{equation}\label{eq12}
  S(x)=\sum_{j=1}^n [ \alpha x_j^2 + (x_{j+1}-x_j )^2 - \chi \exp
  (-i x_j ) ]
\end{equation}
which can be viewed as a lattice field theory for the
one-dimensional classical Yukawa fluid in the grand canonical
ensemble ($\alpha$ is a measure of interaction strength and $\chi$
is the activity). For the case of $n>1$, periodic boundary
conditions are applied. The model has a saddle point $z_j^* = i
y^*_j$ that lies on the imaginary axis and is homogeneous in the
index $j$ (as well as an inhomogeneous ``1d crystal-like'' saddle
point). Its location is given by the solution of $\chi \exp (y^*_j
) + 2 \alpha y^*_j =0$. The optimal displaced path $\bar{y}_j$ is
homogeneous in $j$ and is given by the solution of $\chi \exp
(\bar{y}_j )\langle \cos x_j \rangle_{\bar{y}}+2 \alpha \bar{y}_j
=0$. We see that $y^*$ and $\bar{y}$ are coincident under
conditions ($\alpha \gg 1$) where the random variable $x$
fluctuates closely about the saddle point $x^* =0$. In the
strongly fluctuating regime ($\alpha \ll 1$), $\langle \cos x_j
\rangle_{\bar{y}}$ will be dramatically reduced, resulting in a
large shift of $\bar{y}$ away from $y^*$. These expectations are
borne out in numerical simulations of the model.

\begin{figure}
  \begin{center}
    \epsfig{file=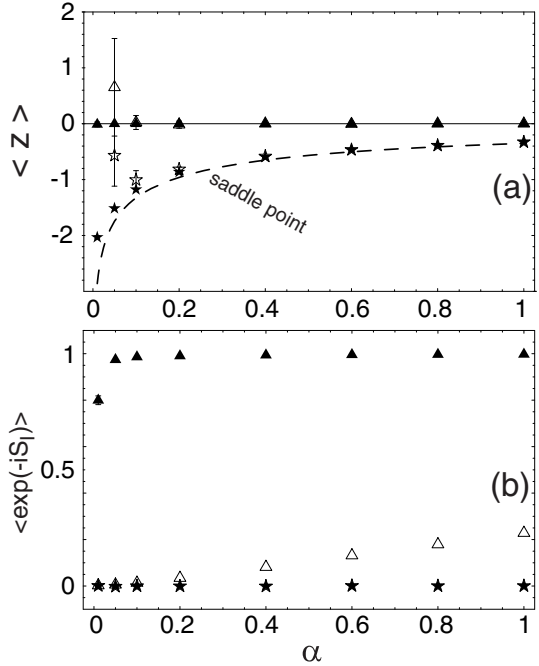,width=7.2cm}
  \end{center}
\caption{Comparison OPS and conventional MC.
  (a): The average of $z=x+i y$ as a function of the
  parameter $\alpha$ for the model of Eq.~(\ref{eq12}) with $n=10$
  and $\chi=1$.  Open and filled symbols are results,
  respectively, from MC and OPS. Stars, triangles, and dashed curve as in Fig.~\ref{fig:fig2}.
  (b): The average sign $\exp (-i S_I )$ for the same parameters
  as in (a). At
  small $\alpha$ the real part of the sign in MC rapidly approaches zero, and the averages
  fail to converge.
  Error bars are comparable to the symbol sizes if not explicitly
  shown.}
\label{fig:fig3}
\end{figure}

We have carried out conventional Metropolis Monte Carlo (MC)[i.e.
Eq.~(\ref{eq5}) with $y=0$], OPS, and CL simulations of the model with
action Eq.~(\ref{eq12}). The results were obtained from runs with a
total of $10^7$ MC cycles or Langevin steps, a time step of 0.001 in
the case of CL, and parameters $M=1000$, $\lambda =0.05$ for OPS. In
Fig.~\ref{fig:fig2} we compare the results obtained from OPS and CL
simulations with $n=1$ and $\chi = 1$. The top panel (a) shows
$\langle z \rangle$ as a function of $\alpha$, while the bottom panel
(b) displays the real and imaginary parts of the ``sign'' $\langle
\exp (-i S_I )\rangle$. In contrast to OPS, CL fails to converge, or
converges very slowly, for $\alpha \lesssim 0.15$. Conventional MC
also converges, but the average sign is approximately 0.8, as opposed
to $\sim 1$ shown by the OPS.

It is often observed\cite{loh} that the sign in conventional MC
simulations decreases exponentially with $n$, causing a breakdown
of the method. This is illustrated for the present model in
Fig.~\ref{fig:fig3} with parameters $n=10$ and $\chi=1$. The
conventional MC method fails to converge for $\alpha\lesssim 0.1$
in contrast to OPS. Moreover, the real part of the sign is
strongly suppressed in the MC results, even at large values of
$\alpha$. The sign problem is evidently strongly suppressed, if
not eliminated entirely for this model in OPS.

The OPS method is applicable to any field theory with an action
$S(z)$ that is analytic throughout a domain of $z$ relevant to
numerical simulations. This includes the important cases of
classical fluids in the grand canonical ensemble and path integral
formulations of time-dependent quantum chemical problems. Other
situations including fluids in the canonical ensemble, strongly
correlated electrons, and lattice gauge theories are characterized
by analytic $\exp (-S)$, but with zeros along the real axis and
hence logarithmic singularities in $S$. We believe that OPS will
also be useful in such problems, however precautions should be
taken to avoid crossing branch cuts in the steepest descent
approach to the optimal displacement $\bar{y}$. Finally, we note
that the displaced paths considered here were parallel to the real
axis. Generalization of the method to optimize both the
displacement and shape of the path could prove even more powerful.

In summary, we have identified a variational principle that
permits a global stationary phase analysis of integrals of
arbitrary dimension with analytic integrands. We expect that this
technique will have important implications for analytical and
numerical investigations of field theories in the complex plane.

\begin{acknowledgments}
This work was supported in part by the NSF under the MRSEC program
award No.\ DMR00-80034 and DMR98-70785. We are grateful to H.
Metiu, C. Garcia-Cervera, R. Sugar, J. S. Langer, M. P. A. Fisher,
and D. Scalapino for helpful discussions.
\end{acknowledgments}

%\bibliography{sign_pb_03}

\begin{thebibliography}{10}
  
\bibitem{lgt} I. Montvay and G. M{\"u}nster, {\em Quantum Fields on the
    Lattice} (Cambridge University Press, Cambridge, 1994).
  
\bibitem{quantum1} V.~S. Filinov, Nuclear Physics B {\bf 271}, 717
  (1986); J.~D. Doll and D.~L. Freedman, Adv. Chem. Phys. {\bf 73},
  289 (1988); N. Makri and W.~H. Miller, Chem. Phys. Lett. {\bf 139},
  10 (1987).

%\bibitem{quantum2}
%J.~D. Doll and D.~L. Freedman, Adv. Chem. Phys. {\bf 73},  289  (1988).

%\bibitem{quantum3}
%N. Makri and W.~H. Miller, Chem. Phys. Lett. {\bf 139},  10  (1987).
  
\bibitem{loh} E.~Y. Loh~Jr. {\it et~al.}, Physical Review B {\bf 41},
  9301 (1990).

\bibitem{ghfrev} G.~H. Fredrickson, V. Ganesan, and F. Drolet,
  Macromolecules {\bf 35}, 16 (2002); S.~A. Baeurle, Phys. Rev. Lett.
  {\bf 89}, 080602 (2002).

%\bibitem{stephan}
%S.~A. B{\"a}urle, Phys. Rev. Lett. {\bf 89},  080602  (2002).

\bibitem{binder}
D.~P. Landau and K. Binder, {\em A Guide to Monte Carlo Simulations in
  Statistical Physics} (Cambridge University Press, New York, 2000).

\bibitem{lin}
H.~Q. Lin and J.~E. Hirsch, Phys. Rev. B {\bf 34},  1964  (1986).

\bibitem{parisi} G. Parisi, Phys. Lett. B {\bf 131}, 393 (1983); J.~R.
  Klauder, Phys. Rev. A {\bf 29}, 2036 (1984).

%\bibitem{klauder}
%J.~R. Klauder, Phys. Rev. A {\bf 29},  2036  (1984).

\bibitem{lee}
S. Lee, Nuclear Physics B {\bf 413},  827  (1994).

\bibitem{schoenmaker}
W.~J. Schoenmaker, Physical Review D {\bf 36},  1859  (1987).

\bibitem{bender}
C.~M. Bender and S.~A. Orszag, {\em Advanced Mathematical Methods for
  Scientists and Engineers} (McGraw-Hill Publishing Company, New York, 1978).

\bibitem{kennedy} A.~D. Kennedy, Parallel Computing {\bf 25}, 1311
  (1999); P.~J. Rossky and J.~D. Doll, J. Chemical Physics {\bf 69},
  4628 (1978).

%\bibitem{rossky}
%P.~J. Rossky and J.~D. Doll, J. Chemical Physics {\bf 69},  4628  (1978).

\bibitem{sabo}
D. Sabo, J.~D. Doll, and D.~L. Freedman, J. Chemical Physics {\bf 116},  3509
  (2002).

\end{thebibliography}

\end{document}